\documentclass[lettersize,journal]{IEEEtran}
\usepackage{amsmath,amsfonts}
\usepackage{algorithmic}
\usepackage{algorithm}
\usepackage{array}
\usepackage[caption=false,font=normalsize,labelfont=sf,textfont=sf]{subfig}
\usepackage{textcomp}
\usepackage{stfloats}
\usepackage{url}
\usepackage{verbatim}
\usepackage{graphicx}
\usepackage{cite}

\usepackage{hyperref}
\usepackage{orcidlink}
\hypersetup{hidelinks}
\usepackage{multirow}
\usepackage{bbding}
\usepackage{pifont}
\usepackage{rotating}
\usepackage{threeparttable}
\usepackage{booktabs}
\usepackage{listings}
\usepackage{wasysym}
\usepackage{xcolor}
\definecolor{codegreen}{RGB}{11, 115, 27}
\definecolor{codegray}{rgb}{0.5,0.5,0.5}
\definecolor{codered}{RGB}{213, 48, 73}
\definecolor{codeblue}{RGB}{44, 44, 255}

\begin{document}

\bstctlcite{IEEEexample:BSTcontrol}

\title{Yama: Precise Opcode-based Data Flow Analysis for Detecting PHP Applications Vulnerabilities}

\author{Jiazhen Zhao$^{\orcidlink{0000-0001-5024-1679}}$, Kailong Zhu$^{\orcidlink{0000-0001-5241-0157}}$, Lu Yu$^{\orcidlink{0000-0002-8524-0139}}$, Hui Huang$^{\orcidlink{0000-0001-7615-9585}}$, Yuliang Lu$^{\orcidlink{0000-0002-8502-9907}}$
\thanks{This work was supported in part by the National Key R\&D Program of China under Grant 2021YFB3100500 and in part by NSFC under Grant 62202484. (Corresponding authors: Kailong Zhu; Yuliang Lu.)}


\thanks{Jiazhen Zhao, Kailong Zhu, Lu Yu, Hui Huang, and Yuliang Lu are with the College of Electronic Engineering, National University of Defense Technology, Hefei 230037, China and also with Anhui Province Key Laboratory of Cyberspace Security Situation Awareness and Evaluation, Heifei 230037, China(e-mail: zhukailong@nudt.edu.cn; luyuliang@nudt.edu.cn)}}

\markboth{\tiny This work has been submitted to the IEEE for possible publication. Copyright may be transferred without notice, after which this version may no longer be accessible.}%
{Shell \MakeLowercase{\textit{et al.}}: A Sample Article Using IEEEtran.cls for IEEE Journals}



\maketitle

\begin{abstract}
Web applications encompass various aspects of daily life, including online shopping, e-learning, and internet banking. Once there is a vulnerability, it can cause severe societal and economic damage. Due to its ease of use, PHP has become the preferred server-side programming language for web applications, making PHP applications a primary target for attackers. Data flow analysis is widely used for vulnerability detection before deploying web applications because of its efficiency. However, the high complexity of the PHP language makes it difficult to achieve precise data flow analysis, resulting in higher rates of false positives and false negatives in vulnerability detection.

In this paper, we present Yama, a context-sensitive and path-sensitive interprocedural data flow analysis method for PHP, designed to detect taint-style vulnerabilities in PHP applications. We have found that the precise semantics and clear control flow of PHP opcodes enable data flow analysis to be more precise and efficient. Leveraging this observation, we established parsing rules for PHP opcodes and implemented a precise understanding of PHP program semantics in Yama. This enables Yama to precisely address the high complexity of the PHP language, including type inference, dynamic features, and built-in functions.

We evaluated Yama from three dimensions: basic data flow analysis capabilities, complex semantic analysis capabilities, and the ability to discover vulnerabilities in real-world applications, demonstrating Yama's advancement in vulnerability detection. Specifically, Yama possesses context-sensitive and path-sensitive interprocedural analysis capabilities, achieving a 99.1\% true positive rate in complex semantic analysis experiments related to type inference, dynamic features, and built-in functions. It discovered and reported 38 zero-day vulnerabilities across 24 projects on GitHub with over 1,000 stars each, assigning 34 new CVE IDs. We have released the source code of the prototype implementation and the parsing rules for PHP opcodes to facilitate future research.

\end{abstract}

\begin{IEEEkeywords}
Web Security, PHP Applications Vulnerability, Data flow Analysis, Opcode-Based.
\end{IEEEkeywords}

\section{Introduction}
\IEEEPARstart{W}{eb} applications are the primary means of delivering information and services on the internet today. As of July 2024, there are 1,104,170,084 websites globally\cite{netcraft}, covering various aspects of daily life, including social media, shopping, online education, internet banking, and remote work websites. A recent survey indicated that PHP is the most popular server-side language in the world, used by more than 76\% of websites\cite{w3techs}. However, PHP applications have a large number of security vulnerabilities. As of July 2024, PHP vulnerabilities accounted for 45.3\% of all entries in the exploit database, exploit-db\cite{exploit-db}. In the real world, a cyberattack occurs approximately every 39 seconds, with around 88.5 million people falling victim to cybercrime annually, causing global companies to suffer losses exceeding \$9.5 trillion\cite{cobalt,explodingtopics}.

Due to the numerous and harmful vulnerabilities possessed by PHP applications, actively adopting and applying effective defense techniques is a key measure to mitigate potential threats, protect user data. Many methods have been proposed to detect PHP application vulnerabilities, which can generally be categorized as dynamic and static approaches. Dynamic methods detect vulnerabilities by simulating attacks, injecting attack payloads into a deployed application, and inspecting the application's output\cite{Eriksson2021,Trickel2023,Drakonakis2023,AlWahaibi2023}, which is conducted in a real environment, resulting in minimal false positives. However, these methods often suffer from limited code coverage due to the dynamic features and complex navigation of PHP applications, leading to potential false negatives\cite{Alhuzali2018}. Additionally, the need to adjust attack payloads results in significant performance overhead for large applications. On the other hand, static analysis methods are known for their ability to comprehensively cover all the code paths of an application, enabling in-depth analysis of every line of code to ensure no omissions\cite{Shi2024,Su2023,Muralee2023,Marashdih2023,Backes2017}. With high code coverage, low-performance overhead, and a high degree of automation, static analysis is a promising approach for vulnerability detection in PHP applications. Researchers are focusing on improving current static analysis methods to discover more security vulnerabilities\cite{wap,Luo2022,Dahse2014,progpilot,Jovanovic2006,Nunes2015,AlKassar2023,Alhuzali2018}.

After a thorough study of existing static vulnerability analysis solutions for PHP applications, we identified the common challenges related to high false positive and false negative rates, arising from three core limitations. Firstly, the data flow analysis implemented in prior work has often been inaccurate, the lack of type declarations and automatic type conversions in PHP complicate type inference and result in semantic loss and incorrect data flow and control flow\cite{Seidel2023}. Secondly, existing approaches methods inadequately handle PHP's dynamic features, such as variable variables\cite{varvar} and dynamic includes, which are common in PHP applications\cite{Hills2013}. This oversight causes data flow analysis to end prematurely, increasing false negatives and reducing analysis accuracy. Thirdly, existing works simplify modeling built-in function semantics, leading to imprecise taint analysis. Over-tainting and early termination are common issues\cite{Jovanovic2006, Nunes2015, progpilot}, as many tools inadequately model encoding/decoding functions or limit the number of functions analyzed, resulting in false positives and negatives in vulnerability detection.

Our goal is to address the limitations of the existing work mentioned above and propose a precise data flow analysis method for PHP applications that can effectively detect vulnerabilities in PHP applications. To achieve this, we conducted an in-depth study of PHP's low-level instruction set, opcodes\cite{nikic}, and performed a comprehensive analysis of the program semantics represented by most of these opcodes. PHP opcodes are presented as three-address codes in an intermediate representation, which ensures highly precise execution semantics\cite{Moeller2023}. This representation allows complex control structures, such as loops and conditionals, to be broken down into simpler jump and comparison operations, making the control flow and execution order of the program more intuitive and clear. The precise expression of program semantics through PHP opcodes enhanced the precision of our data flow analysis.

Our approach operates in two steps. In the first step, Yama constructs control flow graphs (CFGs) for each file based on the jump instructions in the given opcodes. The second step involves performing forward data flow analysis using the CFGs of each file. During the data flow analysis, we infer types based on the semantics of the opcodes. Additionally, we conducted a comprehensive analysis of the opcodes related to file inclusion, function calls, and method calls, which are critical for obtaining accurate interprocedural context. To address the challenges posed by PHP’s extensive dynamic features, we categorized these features into seven types and leveraged the three-address code format of opcodes to decompose the complex semantics of the dynamic features into multiple simpler instructions. Furthermore, we invested substantial effort into comprehensively modeling PHP’s built-in functions and integrating concrete execution to mitigate the challenges posed by built-in functions that cannot be fully analyzed statically.

We implemented a prototype of Yama using over 10K lines of PHP code and thoroughly evaluated it across three dimensions: basic data flow analysis capabilities, complex semantic analysis capabilities, and the ability to discover vulnerabilities in real-world applications. The results indicated that Yama possesses context-sensitive and path-sensitive interprocedural analysis capabilities, achieving a 99.1\% true positive rate in complex semantic analysis experiments related to type inference, dynamic features, and built-in functions. Yama’s vulnerability detection capabilities were evaluated on 24 popular PHP applications having a total of 10M LOCs. Yama was able to complete the analysis of these applications in approximately 30 minutes and detected 109 true positive vulnerabilities, including 38 previously unknown vulnerabilities. A comparison with state-of-the-art tools further highlights Yama’s ability to identify all vulnerabilities reported by other tools, with faster analysis speeds and the highest precision, while also discovering 37 additional vulnerabilities that other tools missed. We conducted an exploitability analysis on these newly discovered vulnerabilities and found that they could indeed be exploited, leading to serious security consequences. We have responsibly disclosed these new vulnerabilities to the developers of the affected applications, and as of the writing of this paper, 38 of the vulnerabilities have been confirmed or patched, including 34 CVEs\cite{Zhao2024}.

In summary, this paper makes the following contributions:

• To the best of our knowledge, we are the first to systematically analyze and summarize the program semantics represented by PHP opcodes. This lays the foundation for further research in the field of PHP program semantics understanding and inspires new directions for exploration.

• We present Yama, a context-sensitive and path-sensitive interprocedural data flow analysis method for PHP taint-style vulnerability detection, improving detection accuracy.

• Yama outperforms the state-of-the-art tools, effectively addressing the challenges posed by PHP's dynamic features and built-in functions. It has discovered 38 previously unknown vulnerabilities, including 34 CVEs.

• The source code of Yama is publicly available at https://github.com/xjzzzxx/Yama, facilitating future research.

\section{Background}

\subsection{PHP program analysis}

Unlike typical desktop applications, PHP, which was designed for web development, has multiple entry points\cite{Backes2017}. This complicates static analysis, as it requires strategies to handle multiple entry points, prioritize them, and meet the data flows. Contextual dependencies also vary, affecting program states. To address these challenges,  it is necessary to use precise analysis methods for PHP applications, such as context-sensitive and path-sensitive analyses.

\subsection{Opcodes}

When a PHP script executed, it's compiled into low-level instruction sequences called PHP \textbf{OPCODES}, executed by the ZendVM. Each instruction, known as an \textbf{Opline}, uses a three-address code format with an \textbf{OPCODE} type, two input operands (\textbf{OP1} and \textbf{OP2}), an output operand (\textbf{RES}), and an \textbf{extended\_value} field for additional modifiers\cite{nikic}. Multiple instruction types are referred to as \textbf{opcodes}.

Most static analysis methods for PHP focus on the abstract syntax tree (AST)\cite{Luo2022,Alhuzali2018,Nunes2015,wap}, which represents code structure but not execution flow. This can lead to control flow errors, such as branches and loops being implicit in AST nodes\cite{Alfred2007}, in data flow analysis. While AST is suitable for flow-insensitive and type analyses, it struggles with more precise analyses needed for vulnerability detection\cite{Moeller2023}. In contrast, PHP opcodes explicitly represent control flows, allowing for accurate construction of control flow graphs and supporting higher-precision data flow analysis.

\section{Motivation and Challenge}

\subsection{Motivation}

Listing \ref{motivationCode} shows a PHP code snippet inspired by an cross-site scripting (XSS) vulnerability in DVWA. This vulnerability exists because the attacker controls the \$\_GET variable at line 14, and without being properly sanitized, its value propagated on lines 9 and 11, and reached the echo statement on line 18 after the function call on line 11.

\begin{lstlisting}[label={motivationCode}, caption = Example of an XSS vulnerability,captionpos=b,float]
<?php
define('ROOT', '../../');
require_once ROOT.'dvwa/includes/dvwaPage.inc.php';
switch($_COOKIE['security']) {
  case 'low': $vulFile = 'low.php';break;
  case 'impossible': $vulFile = 'impossible.php';break;
}
require_once ROOT."vulnerabilities/xss_r/source/{$vulFile}";
$page['body'] .= "</form>{$html}</div>";
$func = "dvwaHtmlEcho";
$func($page);
// vulnerabilities/xss_r/source/low.php
if (array_key_exists("name", $_GET) && $_GET['name'] != NULL) {
  $html .= '<pre>Hello'.$_GET['name'].'</pre>';
}
// dvwa/includes/dvwaPage.inc.php
function dvwaHtmlEcho($pPage){
  echo "... {$pPage['body']}..."; 
}
\end{lstlisting}

To detect this vulnerability, the static application security testing(SAST) tools should first analyze the dynamic include on line 8, which requires correctly handling the switch-branch on line 4 and solving for the value of \$vulFile. Secondly, analyze the if-branch on line 13 and mark the variable \$html on line 14 as a tainted variable, which needs to satisfy the constraints at line 13. Therefore, it is necessary to model the semantics of the built-in function array\_key\_exists correctly. Thirdly, track the propagation of tainted variables on line 9, which involves the propagation of taint from scalar variables (\$html) to an element in the array (\$page ['body']). Finally, analyze the variable function call on line 11 and track the taint until the echo statement on line 18 outputs the tainted variable \$pPage ['body'], which involves variable function inference, data flow between procedures, and the definition of the sink function.

The current state of art SAST tools are unable to detect this vulnerability. RIPS\cite{Dahse2014} is unable to cope with the control flow challenge brought by the switch statement, and the \$vulFile is set to impossibile.php, causing it to erroneously include the file impossibile.php on line 8, which is without vulnerabilities. TCHECKER\cite{Luo2022} is unable to correctly handle the interprocess data flow from file inclusion. It can analyze the source of the taint in low.php file but cannot return the taint. Progpilot\cite{progpilot} can overcome the challenges encountered by the above two tools, but it cannot correctly handle the variable function(\$func) on line 11, as it cannot infer the function body corresponding to the variable function.

\subsection{Challenge}
Our motivation example has prompted us to reflect on the data flow analysis capabilities of current SAST tools, which is the motivation for our research. Actually, PHP is a weakly typed programming language with many dynamic features and built-in functions and methods\cite{Hills2013,Dahse2014}. These unique features of PHP also bring more complex semantics that need to be processed for PHP's data flow analysis. Below, we summarize three challenges that PHP applications face in data flow analysis.

\subsubsection{Inferring Variable Type}
As pointed out by prior work, variable inference is essential to data flow analysis\cite{Fink2000,Seidel2023}. Suppose the variable type is incorrectly determined during the data flow analysis of a PHP application. This may result in semantic loss and lead to incorrect information about the data flow and control flow related to that variable. For example, treating an array variable as a scalar variable can result in semantic loss of each element in the array, leading to errors when parsing semantics such as array loops and element reads.

Listing \ref{varInfer} takes array variables as an example to further introduce the importance of variable type inference. On line 2 of the code, the taint is passed to variable \$b through variable \$\_GET ['p1'], and then passed to the fourth element in the array variable \$array on line 4. At this point, all other elements in the \$array do not carry any taint. After passing the if-branch on line 5, the echo statement on line 6 outputs the fourth element of the \$array, resulting in an XSS vulnerability. If the array variable types are not correctly inferred, analysis errors may occur in the if-branch on line 4 and the tainted variable recognition on line 5. 

\begin{lstlisting}[label={varInfer}, caption = Example of variable type inference,captionpos=b,float]
<?php
$b = $_GET['p1'];
$array = array('abc', 'def', 'ghi');
$array[] = $b;
if($array[1]){
  echo $array[4];
}
\end{lstlisting}

\subsubsection{Dynamic Features} \label{challenge2}

As is well known, dynamic features are challenging to model in static analysis\cite{Hills2013,Hills2015,Kyriakakis2019}. When analyzing the data flows of PHP applications, a difficult problem that cannot be avoided is how to handle the dynamic features. We reclassified PHP's dynamic features into seven types: \textbf{variable variables(D1)}, \textbf{dynamic includes(D2)}, \textbf{dynamic PHP code execution(D3)}, \textbf{variadic functions(D4)}, \textbf{variable function(D5)}, \textbf{variable object(D6)}, \textbf{magic methods(D7)}, whose definitions can be found in other works\cite{Dahse2014,Hills2013}. Listing \ref{df} takes variadic functions as an example to introduce the challenges of modeling each dynamic feature. 

\begin{lstlisting}[label={df}, caption = Example of variadic functions, captionpos=b,float]
<?php
function sum(...$numbers){
  foreach ($numbers as $n) {
    echo $n;
  }
}
$b = $_GET["p1"];
sum(1, 2, 3, $b);
\end{lstlisting}

Firstly, it is declared that the function sum accepts variable arguments by ellipsis(line 2). Then multiple parameters, including the tainted variable \$b, are accepted when the function sum is called (line 8). These parameters are assigned as an array to the parameter variable \$numbers in the function body. Thus, the fourth element of the array \$numbers carries the taint. After iteration, \$n with tainted is output (line 4), resulting in an XSS vulnerability. Since variadic functions can accept any number of parameters, it is difficult to accurately infer the number and type of parameters in static analysis, which affects the inspection and analysis of function calls and brings difficulties to data flow analysis.

\subsubsection{Built-In Functions}

PHP offers many convenient built-in functions and is widely used by developers, which poses an significant challenge for PHP program analysis\cite{Huang2019}. PHP built-in functions are implemented in C language without PHP code entities, making it challenging for the analyzer to delve into the implementation details of these functions\cite{Li2021}. Currently, modeling the built-in function through expert experience is the favored method, but ensuring the accuracy of modeling semantics of built-in functions is challenging. First, PHP has many built-in functions, and it isn't easy to conduct semantic modeling for all built-in functions. Secondly, the built-in functions of PHP may produce different semantics under different parameters, so it is not easy to guarantee the semantic accuracy of manual modeling.

\begin{lstlisting}[label={c3:bif}, caption = Example of built-in function,captionpos=b,float]
<?php
$val = $_GET["p1"];
$input = array(12, 10, 9);
$result = array_pad($input, 5, $val);
echo $result[3];
\end{lstlisting}

Listing \ref{c3:bif} shows an example of a challenge posed by a built-in function for data flow analysis. In this example, the taint is received by \$\_GET on line 2, and then the tainted variable \$val is passed to the array \$result on line 4 through the built-in function array\_pad. This built-in function fills the array \$input with \$val to the length of the second argument's value. At this point, the fourth and fifth elements of \$result are both tainted variables \$val and finally, the fourth element of \$reuslt with the taint is output on line 5, causing an XSS vulnerability. In this example, if the built-in function array\_pad is not modeled correctly, the data flow passing the taint from \$val to \$result will be lost.

\section{Design}

We present Yama, a precise context-sensitive path-sensitive interprocedural PHP data flow analysis method. It can precisely infer variable types, handle the complex dynamic features of PHP, and mitigate the problem of built-in function analysis through concrete execution. The high-level design of Yama is first introduced in subsection \ref{Design:view}, and its two major components are elaborated in subsections \ref{Design:cfg} and \ref{Design:dataflow}.

\subsection{Overview} \label{Design:view}
\begin{figure}[!t]
\centering
\includegraphics[width=3.7in]{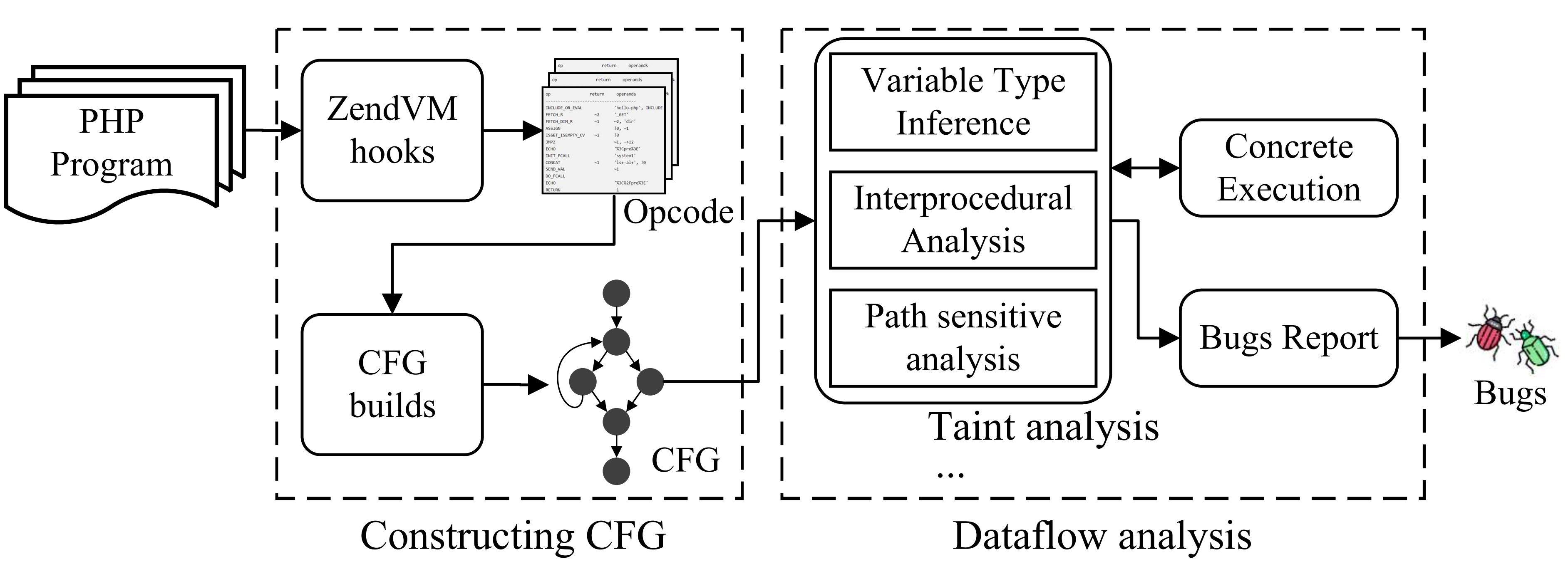}
\caption{The overall architecture of Yama.}
\label{fig:overview}
\end{figure}

As shown in Figure \ref{fig:overview}, Yama has two main steps: (1) Constructing control flow graph and (2) Data flow analysis.

Given the source codes of a PHP application, the first step is to build CFGs of the application's files. At this stage, the opcodes of each application file are obtained via the kernel hook of the ZendVM. Then, the CFG of each file is constructed according to jump instructions in opcodes.

The second step is to perform data flow analysis according to the CFG of each file. Yama has the scalability to carry out a variety of classical data flow analysis methods, such as available expression analysis, reaching definition analysis, and live variable analysis. In this method, we use taint analysis to identify taint-style vulnerabilities\cite{Luo2022} in PHP applications. Based on accurate variable type inference, Yama defined the taint criterion (sources, sanitizers and sinks) and achieved precise context-sensitive and path-sensitive interprocedural taint analysis. To mitigate the challenges of some dynamic features and built-in functions, Yama uses concrete execution to solve for concrete values. After completing the analysis, Yama reports all instances of tainted data flowing to sinks without sufficient sanitization.

\subsection{Constructing CFG} \label{Design:cfg}

PHP applications typically consist of multiple files containing a portion of functionality or code, giving the program numerous entry points. Therefore, each PHP file in the PHP application is regarded as a basic analysis unit that performs the CFG construction for each PHP file. In this stage, the first step is to use the PHP extension, i.e., Vulcan Logic Dumper(VLD)\cite{derickr}, to hook the ZendVM, and obtain the opcodes of the PHP files, then save them to the file whose suffix is .opcode. It is worth mentioning that we have made some changes to VLD, making it output some additional information to help us solve some problems of dynamic features, such as the properties of the function parameter (whether the reference is\_ref and the variable parameter is\_variadic), function declared static variable information, class properties, class inheritance relationship, traits (code reuse), and so on.

In the second step, the opcodes are analyzed, and the CFG of each file is constructed according to the jump instructions. Specifically, we determine our basic block entry point based on three principles: 

a) The first instruction of the instruction sequence is a basic block entry point. 

b) The address specified by the jump instruction is the entry point of the basic block.

c) The next instruction of the jump instruction is the entry point of the basic block.

Note that all instructions between each basic block entry point and the next basic block entry or program end belong to the same basic block.

\subsection{Data flow analysis} \label{Design:dataflow}

Yama designed precise taint analysis(\ref{dataflow:taint}) to identify taint-type vulnerabilities and solved three challenges in PHP static analysis through a series of steps: firstly, systematically summarized the characteristics of opcodes type dependencies to achieve accurate variable type inference(\ref{dataflow:var}); Then, through precise interprocedural analysis(\ref{dataflow:inter}) and path-sensitivity analysis(\ref{dataflow:path}), ensuring the accuracy of the data flow and combining them with the accurate inference of the variable types involved in dynamic features, the challenge of dynamic features is solved; Finally, we obtain the specific return value of the built-in function directly through concrete execution(\ref{dataflow:concr}), mitigating built-in function analysis challenges that cannot be statically processed.

\subsubsection{Taint Analysis} \label{dataflow:taint}

We use taint analysis to track untrusted data flows into critical operations, thereby identifying potential taint-style vulnerabilities in an application. Yama designates five user-controllable superglobals as taint sources(\$\_GET, \$\_POST, \$\_FILES, \$\_COOKIE and \$\_REQUEST) and based on taint propagation rules of PHP opcodes, implement the tracking of interprocedural taint propagation. Due to space limitations, the taint propagation rules of PHP opcodes are provided in our repository\cite{Zhao2024}. Specially, Yama identifies two types of taint sanitization: irreversible, which stops taint propagation, and reversible, which allows taint restoration using a paired function(e.g., htmlspecialchars and htmlspecialchars\_decode) tracked via a stack. A variable is tainted if the stack is empty and taint exists. Taint sink refers to the situation where a tainted variable passes to a dangerous function, potentially leading to vulnerabilities. Not all parameters in these functions are dangerous, so Yama maintains a list of dangerous functions containing the positions of the dangerous parameters(dangerous parameters list) from 8 common taint-style vulnerabilities, including cross-site scripting(XSS), SQL injection(SQLI), remote command execution and code execution(RCE), file inclusion(FI), arbitrary file deletion(AFD), unrestricted file upload(UFU), path traversal(PT) and sensitive data exposure(SDE).

\subsubsection{Variable Type Inference} \label{dataflow:var}
As mentioned earlier, incorrectly identifying variable types in the program can lead to imprecise data flow analysis. Therefore, Yama needs to correctly identify and infer the type of each variable during the analysis process.

Firstly, we categorize PHP variables into three types: scalar\cite{PHPscalar}, array, and object. After our in-depth research on PHP opcodes, we found that the opcodes are type-dependent. For example, the instructions INIT\_ARRAY and ADD\_ARRAY\_ELEMENT are specific to array types, while NEW and ASSIGN\_OBJ are specific to object types. Based on this discovery, we can infer types according to different instructions. Due to space limitations, the dependent types for each opcode are provided in our repository\cite{Zhao2024}. Below are methods for type inference of scalars, arrays, and objects. Scalars are the foundation of variable types, including int, float, strings, and boolean types. They represent a single value, the smallest unit of data types. The combination of multiple scalars can form properties in arrays or objects. There are two main situations for inferring scalars: first, when encountering instructions such as assignment, if the R-value is a constant, it means that the L-value must be a scalar. Another common situation is that when encountering a variable as an operand, we use a function called \textit{varClassify} to determine the variable type and then retrieve the data structure based on its type. For arrays and objects, type inference is mainly based on their specific instructions.

\begin{figure}[!t]
\centering
\includegraphics[width=3.5in]{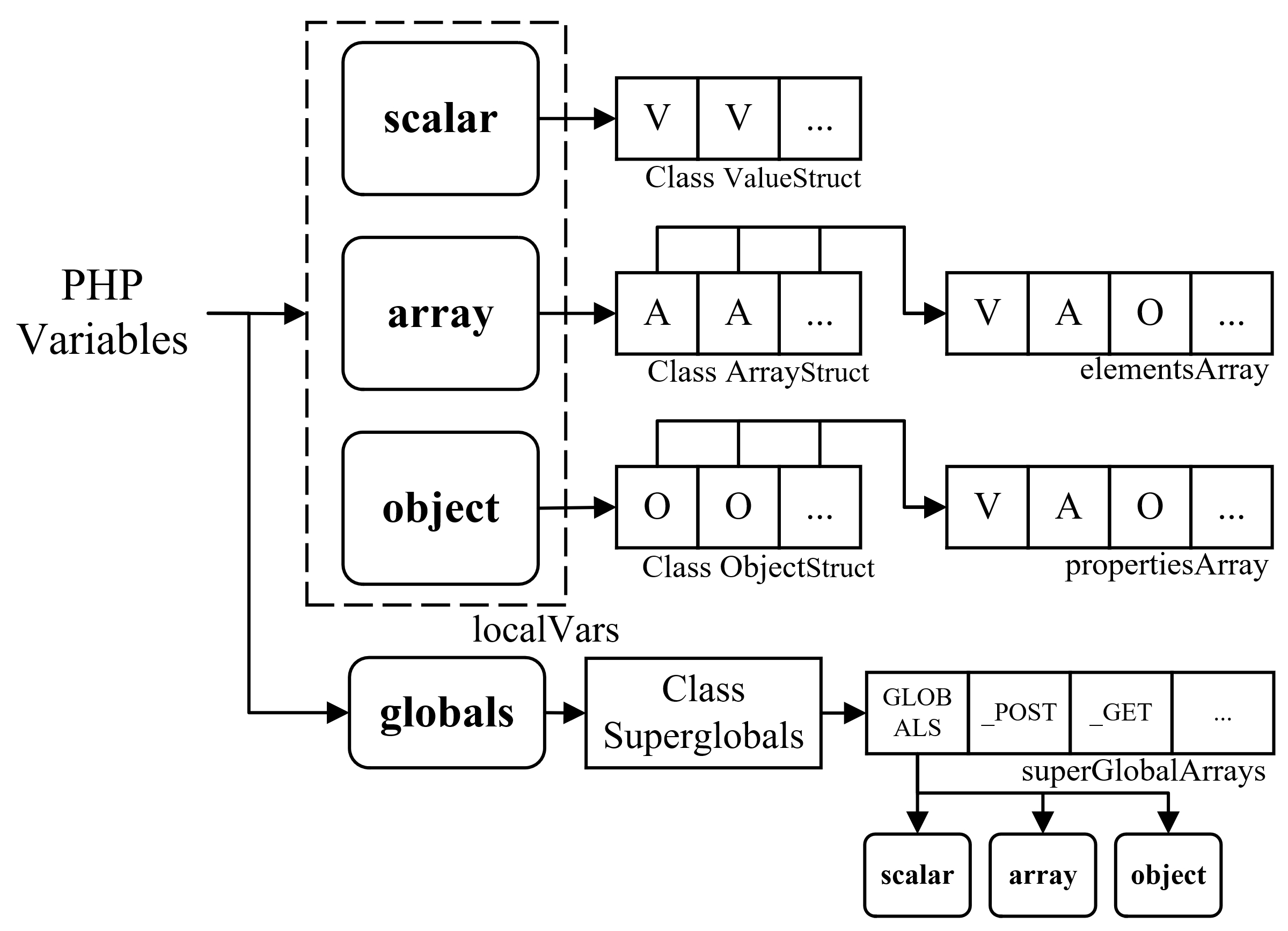}
\caption{The data storage model of Yama.}
\label{fig:typeinfer}
\end{figure}

In order to improve the efficiency of type inference, we designed a data storage model to classify and store variable data of scalar, array, object, and global variables, as shown in Figure \ref{fig:typeinfer}. Specifically, the global here store references to global variables in the current scope and nine types of superglobals. After correctly inferring the type of each variable during the analysis, Yama instantiated the corresponding object and stored information, such as \textit{ValueStruct}, \textit{ArrayStruct}, and \textit{ObjectStruct}.

\subsubsection{Interprocedural Analysis} \label{dataflow:inter}
Yama performs context-sensitive analysis for the two interprocedural scenarios in PHP: file inclusion and function calls (including method calls).

\textbf{File Inclusion}. PHP implements file inclusion through the INCLUDE\_OR\_EVAL instruction, where the path of the included file is specified by an operand. For clarity in our discussion, we refer to the included file as \textbf{INC}. In practice, we have identified that the paths to INC can be classified into four cases\cite{PHPinclude}, as shown in Listing \ref{fileinclude}.

The first is the typical case, where the path value is resolved through string concatenation. Yama needs to model the define statements and the array \$a, then concatenate them on line 5 of the code to derive the path of INC. The second case is include\_path, an environment variable in PHP. When the file specified by the INC path cannot be found, PHP treats the INC path as a relative path and searches for the file within the directories specified by include\_path. On line 7 of the code, an additional include\_path is added via two built-in functions; therefore, Yama must model these two built-in functions to locate INC accurately. The third case pertains to the working directory. When the first two methods fail to locate INC, Yama relies on the current working directory to search for the file. On line 10 of the code, the built-in function chdir changes the current working directory. Consequently, Yama needs to model chdir and maintain the value of the current working directory. The fourth case involves dynamic INC paths. In this case, Yama reports a file inclusion vulnerability because the included file can be specified by the user.

\begin{lstlisting}[label={fileinclude}, caption = Example of file inclusions,captionpos=b,float]
<?php
# case 1: typical
define('DIR_ROOT', 'www/');
$a = array('upload/','images/');
include(DIR_ROOT . $a[0] .'a.php');
# case 2: include_path
set_include_path( get_include_path() . 'lib/' );
include('Model/Init.php');
# case 3: working directory
chdir("config/"); 
include("db.php");
# case 4: dynmaic
include($_GET['filepath']);
\end{lstlisting}

\textbf{Function Calls}. We primarily focus on the data flow within the called function's process and the data flow along the function call and return edges. PHP function calls involve five types of instructions: 

a) Initialization, Executed through instructions resembling INIT*CALL*, which locate the function body based on the function name and method name provided by the operands. Leveraging Yama's accurate type inference and variable storage, the corresponding function body of the called function can be accurately identified, thereby enabling correct analysis of the data flow within the called function. 

b) Parameter Passing, Specified by instructions in the form of SEND\_*, designates the function's formal parameters. 

c) Function Execution, Performed via DO\_FCALL instructions, where the return value of the function execution is received through the output operand RES. 

d) Parameter Receiving, Within the body of the called function, parameter information is obtained through instructions such as RECV\_*. 

e) Function Return, The called function's body uses opcodes like *RETURN* to provide the function's return value.

The precise parsing of function parameter passing and return value handling allows Yama to accurately track the data flow along the call and return edges.

\subsubsection{Path-Sensitive Analysis} \label{dataflow:path}
Path-sensitive analysis computes different pieces of analysis information dependent on the predicates at the conditional branch, which makes data flow analysis more precise. When encountering conditional jump instructions, Yama can compute these conditions such as numerical comparisons (e.g., a==b), logical judgments (e.g., \&\&), variable settings (e.g., isset), and so on. However, Yama faces challenges because some conditions can only be computed at runtime. In such cases, Yama assumes that all branch paths may be executed. Listing \ref{pathScode} presents an interesting example where flow-sensitive analysis considers there is an XSS vulnerability in the codes, while path-sensitive analysis does not think so. Yama executes 10 iterations based on the condition $\$i<10$. Since it is impossible for \$i to equal 20, Yama does not mistakenly report a vulnerability in this case.

\begin{lstlisting}[label={pathScode}, caption = Example of a safe code,captionpos=b,float]
<?php
$x = 10; $y = 0;
for ($i=0; $i < 10; $i++) { 
    if($i == 20){
        $z = $_POST['xss']; echo $z;
    }
}
\end{lstlisting}

\subsubsection{Concrete Execution} \label{dataflow:concr}

In Yama, we use concrete execution to mitigate built-in function analysis challenges that cannot be statically processed. Yama maintains a specific list of built-in functions to be concretely executed. When the called function is contained in this list, the built-in function is executed, and the function's return value is saved in the RES operand of the DO\_CALL instruction. Yama implements concrete execution for two types of built-in functions. The first type is that the function's return value cannot be obtained through static analysis; for example, the return value of get\_include\_path is an environment variable. The second type is built-in functions that are difficult to model semantically, such as parse\_str, which parses a string to multiple variables.

\section{Evaluation}
In our research, we have found three primary reasons why many static analysis tools struggle to detect vulnerabilities effectively. First, these tools exhibit deficiencies in their implementation of data flow analysis, such as the lack of interprocedural analysis capabilities, which results in incomplete and inaccurate program state information derived from data flow analysis. Second, the high complexity of PHP, including type inference, dynamic features, and built-in functions, complicates static analysis. Third, real-world applications often feature robust modular designs, with extensive use of file inclusions and function calls, which elongate data flow paths and consequently reduce the precision of data flow analysis.

Therefore, we evaluate Yama's performance from three dimensions: basic data flow analysis capabilities, complex semantic analysis capabilities, and the ability to identify vulnerabilities in real-world applications. We benchmark Yama against state-of-the-art tools. The evaluation focuses on addressing the following research questions (RQs):

\textbf{RQ1}: Does Yama possess basic data flow analysis capabilities, such as context sensitivity, path sensitivity, and interprocedural analysis?

\textbf{RQ2}: Can Yama effectively address the three current challenges in PHP static analysis?

\textbf{RQ3}: How does Yama perform in analyzing large-scale real-world applications?

\subsection{Datasets and Experimental Setup}

\textbf{Dataset 1} (Expert Experience): Current evaluations of static analysis tools' data flow capabilities are mostly qualitative. To improve this, we constructed a dataset based on expert experience to assess tools' abilities in path-sensitive, context-sensitive, and interprocedural analysis. In this way, we can determine their data flow analysis capabilities more accurately. Details on the evaluation principles are in Section \ref{eva:rq1}.

\textbf{Dataset 2} (Testability Tarpits): Testability Tarpits\cite{AlKassar2022} are code patterns that challenge static analysis tools in detecting vulnerabilities. Al Kassar\cite{AlKassar2022} studied these patterns in PHP and JavaScript, providing a benchmark suite. We used the PHP Testability Tarpits\cite{enferas2021a} from this suite to evaluate the semantic analysis capabilities of PHP static analysis tools, correcting errors and categorizing them into three types based on analysis challenges, as detailed in Table \ref{yama_table_dataset2_ov}.

\begin{table}[htbp]
	\centering
	\caption{Overview of Dataset 2}
	\begin{tabular}{cccc}
		\hline
		\textbf{Type}  & \textbf{Vul} & \textbf{Neg} & \textbf{Total}\\
		\hline
		C1\_type\_inference & 52    & 8     & 60 \\
		C2\_Dynamic\_features & 30    & 1     & 31 \\
		C3\_Built-in\_functions & 29    & 3     & 32 \\
		\hline
		SUM & 111   & 12    & 123 \\
		\hline
	\end{tabular}%
	\label{yama_table_dataset2_ov}%
\end{table}%

\textbf{Dataset 3} (Real World Applications): We selected 24 PHP applications, totaling over 10 million lines of code, as our real-world application dataset, as shown in Table \ref{yama_table_dataset3_ov}. These applications were chosen based on two criteria: (1) Popularity: applications with over 1K stars on GitHub, including well-known PHP applications such as WordPress and Joomla. (2) Previously used datasets: applications utilized in related work.

\begin{table}[htbp]
	\centering
	\caption{Overview of Dataset 3}
        \resizebox{0.5\textwidth}{!}{
	\begin{tabular}{ccccc|ccccc}
		\hline
		\textbf{\#}    & \textbf{Application} & \textbf{Version} & \textbf{Stars} & \textbf{LoC}   & \textbf{\#}    & \textbf{Application} & \textbf{Version} & \textbf{Stars} & \textbf{LoC} \\
		\hline
    1     & wordpress & 6.6   & 19.1k & 585,422 & 13    & openemr & 6\_0\_0 & 2.9k  & 844,467 \\
	2     & DVWA  & 1.9   & 9.7K  & 33,884 & 14    & shopware & 6.4.4.0 & 2.7k  & 581,328 \\
	3     & dolibarr & 12.0.0 & 5.1k  & 63,445 & 15    & phpipam & 1.6   & 2.2k  & 194,546 \\
	4     & organizr & 1.9   & 5k    & 1,331,842 & 16    & WDScanner & latest & 2k    & 6,692 \\
	5     & Joomla & 5.1.2 & 4.7k  & 979,753 & 17    & bjyadmin & latest & 1.8k  & 288,839 \\
	6     & SuiteCRM & 7.12.6 & 4.3k  & 968,371 & 18    & Gazelle & latest & 1.8k  & 85,026 \\
	7     & leantime & 2.1.5 & 4.3k  & 1,492,972 & 19    & phpbb & 3.3.10 & 1.8k  & 355,350 \\
	8     & glpi  & 10.0.16 & 4k    & 36,930 & 20    & unmark & 1.9.2 & 1.6k  & 83,409 \\
	9     & dzzoffice & 2.02.1 & 3.9k  & 196,872 & 21    & icecoder & 8.1   & 1.4k  & 14,040 \\
	10    & librenms & 21.1.0 & 3.6k  & 258,484 & 22    & openflights & latest & 1.4k  & 12,049 \\
	11    & microweber & 2.0.16 & 3.1k  & 409,969 & 23    & RPI-Jukebox & 2.7.0 & 1.3k  & 10,128 \\
	12    & microweber & 1.2.3 & 3.1k  & 291,880 & 24    & PicUploader & latest & 1.2k  & 1,516,425 \\
		\hline
	\end{tabular}%
        }
	\label{yama_table_dataset3_ov}%
\end{table}%

\textbf{Baseline}: We selected baselines based on two criteria: they must be open source and executable, and they must be state-of-the-art or commonly used in renowned benchmarks with significant impact. Based on these criteria, we searched recent studies on PHP application vulnerability static analysis. We selected eight state-of-the-art tools: WAP\cite{wap}, Progpilot\cite{progpilot}, Pixy\cite{Jovanovic2006}, RIPS\cite{Dahse2014}, PhpSAFE\cite{Nunes2015}, NAVEX\cite{Alhuzali2018}, TCHECKER\cite{Luo2022}, and WHIP\cite{AlKassar2023}. Although PHPJoern\cite{Backes2017} also met our criteria, it was not selected as TCHECKER and NAVEX already include parts of PHPJoern's code.

\textbf{Experimental Setup}: Experiments are conducted on a machine with an Intel Core i7-10875h processor and 16 GB RAM, using either Windows 10 or Linux (Ubuntu 18.04/22.04). PHP versions 8.0.25 and 7.2.25, JDK versions 1.8\_221 and 11.0.22, and Apache 2.4 are used. To ensure compatibility with different baselines, both older and newer environments are configured, such as Ubuntu 18.04 for Pixy and Ubuntu 22.04 for WHIP.

Resources for the aforementioned datasets and baseline work are available in our repository\cite{Zhao2024}.

\subsection{Data Flow Analysis Capability Assessment (RQ1)} \label{eva:rq1}

In this section, we evaluate the data flow analysis capabilities of Yama and baseline using dataset 1. We further explain how this dataset evaluates the data flow analysis capabilities of each tool and analyze the experimental results for each tool.

\begin{figure}[!t]
\centering
\includegraphics[width=3.5in]{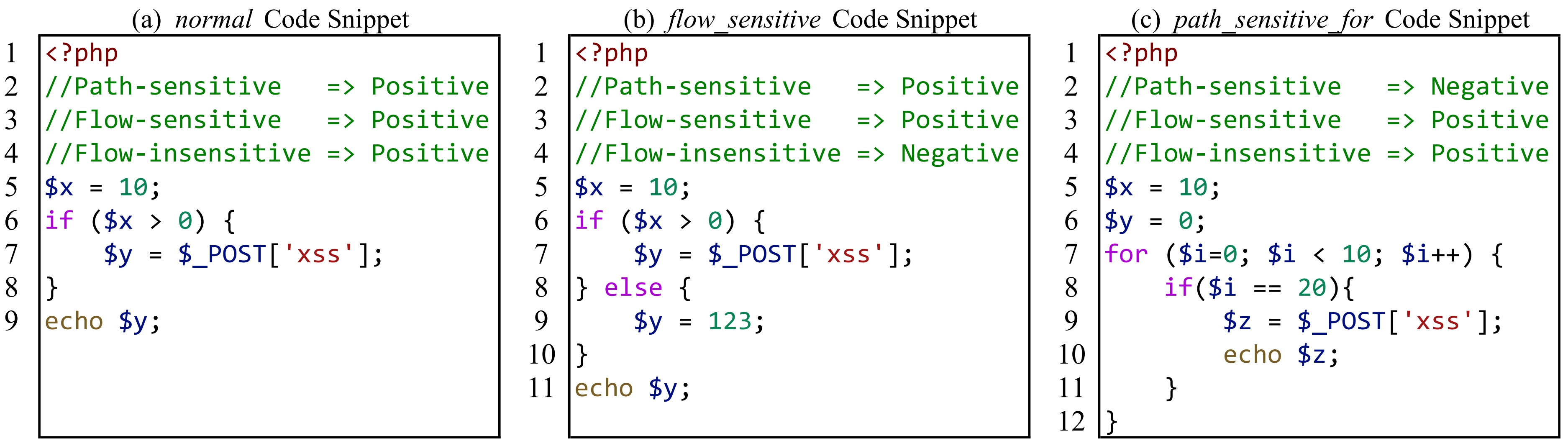}
\caption{Example code from dataset 1.}
\label{fig:yama_code1}
\end{figure}

\textbf{Principles of Dataset 1}. We illustrate the principle of the dataset using the flow-sensitive analysis capability as an example. The sample codes \textit{normal}, \textit{flow\_sensitive}, and \textit{path\_sensitive\_for} are shown in Figure \ref{fig:yama_code1}.
Figure \ref{fig:yama_code1}(b) shows a PHP code snippet with an XSS vulnerability, where a taint is introduced on line 7 and then output on line 11, leading to the XSS vulnerability. The key to detecting this example is having flow-sensitive analysis capability; if the tool is only flow-insensitive, the taint introduced on line 7 would be sanitized during the execution of line 9. Therefore, the flow-sensitive analysis capability of the tool is determined based on the analysis result of this example, as shown in lines 2-4. If the analysis result indicates no vulnerability, the tool is flow-insensitive, and vice versa, it is considered flow-sensitive or path-sensitive. Figure \ref{fig:yama_code1}(c) depicts PHP code without an XSS vulnerability, where the constraint on line 8 cannot be satisfied, preventing the execution of the vulnerable code on lines 9-10. However, both flow-sensitive and flow-insensitive tools mistakenly identify (c) as having an XSS vulnerability, as they do not consider whether constraints are satisfied.

By evaluating the results shown in Figure \ref{fig:yama_code1}(b) and (c), we can determine the tool's flow-sensitive analysis capability. We use $b^T$ and $b^N$ as the cases where the tool classifies (b) as positive and negative, respectively; similarly, $c^T$ and $c^N$ are the classification outcomes for (c). When the analysis result is $b^T \&\& c^N$, it indicates the tool possesses path-sensitive analysis capability; result $b^T \&\& c^T$ indicates the tool has flow-sensitive capability; and the result is $b^N \&\& c^T$ indicates the tool only has flow-insensitive capability.

In reality, Figure \ref{fig:yama_code1}(a) is a redundant code example to ensure the analysis results for (b) and (c) are not coincidental. The codes \textit{normal\_safe} and \textit{path\_sensitive\_if}, which are not shown in Figure \ref{fig:yama_code1}, serve a similar purpose. The context-sensitive analysis capability test cases evaluate both procedural and object-oriented context-sensitive capabilities; the file inclusion analysis capability test cases focus on solving the path of included files, the custom functions within included files, and data flow within included files; finally, the function call analysis capability test cases evaluate whether taints are transmitted through function parameters or returned as function return values. The source code for the above four types of data flow analysis capability test cases is available in our repository\cite{Zhao2024}.

The experimental results are shown in Table \ref{yama_table_rq1}. This table presents the evaluation outcomes of each tool on four types of data flow analysis capability test cases, where $\checked$ indicates a correct evaluation and $\times$ indicates an incorrect one. We use three circular symbols to assess whether each tool possesses a specific data flow analysis capability: $\Circle$ indicates a lack of capability, $\LEFTcircle$ indicates partial capabilities, and $\CIRCLE$ indicates full capabilities. \textbf{It is evident that Yama possesses all four data flow analysis capabilities, receives four $\CIRCLE$, and performs best among all analysis tools}. Following Yama are WHIP, THECKER, Progpilot(P), Pixy, and RIPS, each receiving one $\CIRCLE$ and two $\LEFTcircle$. The remaining tools, ranked by performance from best to worst, are PhpSafe, NAVEX, and WAP.

\begin{table*}[htbp]
	\centering
	\caption{Evaluation of Data flow analysis capability}
	\resizebox{\textwidth}{!}{
		\begin{tabular}{c|cccccc|cccc|cccccccc|ccccccccc}
			\hline
			& \multicolumn{6}{c|}{\begin{tabular}[c]{@{}c@{}}Flow Sensitivities\end{tabular}} & \multicolumn{4}{c|}{\begin{tabular}[c]{@{}c@{}}Context\\ Sensitivities\end{tabular}} & \multicolumn{8}{c|}{\begin{tabular}[c]{@{}c@{}}File Inclusion\end{tabular}}          & \multicolumn{9}{c}{\begin{tabular}[c]{@{}c@{}}Function Call\end{tabular}} \\
			\cline{1-28}
			Tool names & \multicolumn{1}{l}{\begin{sideways}normal\end{sideways}} & \multicolumn{1}{l}{\begin{sideways}normal\_safe$^N$\end{sideways}} & \multicolumn{1}{l}{\begin{sideways}flow\_sensitive\end{sideways}} & \multicolumn{1}{l}{\begin{sideways}path\_sensitive\_if\end{sideways}} & \multicolumn{1}{l|}{\begin{sideways}path\_sensitive\_for\end{sideways}} & \multicolumn{1}{l|}{\begin{sideways}\textbf{ASSESSMENT}\end{sideways}} & \multicolumn{1}{l}{\begin{sideways}normal\end{sideways}} & \multicolumn{1}{l}{\begin{sideways}OO\end{sideways}} & \multicolumn{1}{l|}{\begin{sideways}OO\_safe$^N$\end{sideways}} & \multicolumn{1}{l|}{\begin{sideways}\textbf{ASSESSMENT}\end{sideways}} & \multicolumn{1}{l}{\begin{sideways}normal\end{sideways}} & \multicolumn{1}{l}{\begin{sideways}normal\_safe$^N$\end{sideways}} & \multicolumn{1}{l}{\begin{sideways}func\_normal\end{sideways}} & \multicolumn{1}{l}{\begin{sideways}func\_normal\_safe$^N$\end{sideways}} & \multicolumn{1}{l}{\begin{sideways}concat\end{sideways}} & \multicolumn{1}{l}{\begin{sideways}concat\_safe$^N$\end{sideways}} & \multicolumn{1}{l|}{\begin{sideways}dym\end{sideways}} & \multicolumn{1}{l|}{\begin{sideways}\textbf{ASSESSMENT}\end{sideways}} & \multicolumn{1}{l}{\begin{sideways}normal\end{sideways}} & \multicolumn{1}{l}{\begin{sideways}normal\_safe$^N$\end{sideways}} & \multicolumn{1}{l}{\begin{sideways}AnR\end{sideways}} & \multicolumn{1}{l}{\begin{sideways}AnR\_safe$^N$\end{sideways}} & \multicolumn{1}{l}{\begin{sideways}RnA\end{sideways}} & \multicolumn{1}{l}{\begin{sideways}RnA\_safe$^N$\end{sideways}} & \multicolumn{1}{l}{\begin{sideways}nAnR\end{sideways}} & \multicolumn{1}{l|}{\begin{sideways}nAnR\_safe$^N$\end{sideways}} & \multicolumn{1}{l}{\begin{sideways}\textbf{ASSESSMENT}\end{sideways}} \\
			\cline{1-28}
			WAP   & $\checked$     & $\checked$     & $\times$     & $\times$     & \multicolumn{1}{l|}{$\times$}     & \Circle & $\times$     & $\times$     & \multicolumn{1}{l|}{$\checked$}     & \Circle     & $\times$     & $\checked$     & $\checked$     & $\checked$     & $\times$     & $\checked$     & \multicolumn{1}{l|}{$\times$}     & \Circle & $\checked$     & $\times$     & $\checked$     & $\checked$     & $\checked$     & $\checked$     & $\times$     & \multicolumn{1}{l|}{$\checked$}     & \LEFTcircle \\
			Progpilot(P) & $\checked$     & $\checked$     & $\checked$     & $\times$     & \multicolumn{1}{l|}{$\times$}     & \LEFTcircle  & $\times$     & $\checked$     & \multicolumn{1}{l|}{$\times$}     & \Circle     & $\checked$     & $\checked$     & $\checked$     & $\checked$     & $\checked$     & $\checked$     & \multicolumn{1}{l|}{$\checked$}     & \CIRCLE & $\checked$     & $\times$     & $\checked$     & $\checked$     & $\times$     & $\checked$     & $\checked$     & \multicolumn{1}{l|}{$\checked$}     & \LEFTcircle \\
			Pixy  & $\checked$     & $\times$     & $\checked$     & $\times$     & \multicolumn{1}{l|}{$\times$}     & \LEFTcircle  & $\times$     & $\times$     & \multicolumn{1}{l|}{$\checked$}     & \Circle     & $\checked$     & $\checked$     & $\checked$     & $\checked$     & $\times$     & $\checked$     & \multicolumn{1}{l|}{$\times$}     & \LEFTcircle & $\checked$     & $\checked$     & $\checked$     & $\checked$     & $\checked$     & $\checked$     & $\checked$     & \multicolumn{1}{l|}{$\checked$}     & \CIRCLE \\
			RIPS  & $\checked$     & $\times$     & $\checked$     & $\times$     & \multicolumn{1}{l|}{$\times$}     & \LEFTcircle  & $\times$     & $\checked$     & \multicolumn{1}{l|}{$\times$}     & \Circle     & $\checked$     & $\checked$     & $\checked$     & $\checked$     & $\checked$     & $\checked$     & \multicolumn{1}{l|}{$\times$}     & \LEFTcircle & $\checked$     & $\checked$     & $\checked$     & $\checked$     & $\checked$     & $\checked$     & $\checked$     & \multicolumn{1}{l|}{$\checked$}     & \CIRCLE \\
			PhpSafe & $\checked$     & $\checked$     & $\times$     & $\times$     & \multicolumn{1}{l|}{$\times$}     & \Circle & $\times$     & $\checked$     & \multicolumn{1}{l|}{$\times$}     & \Circle     & $\checked$     & $\checked$     & $\checked$     & $\checked$     & $\times$     & $\checked$     & \multicolumn{1}{l|}{$\times$}     & \LEFTcircle & $\checked$     & $\checked$     & $\checked$     & $\checked$     & $\checked$     & $\checked$     & $\checked$     & \multicolumn{1}{l|}{$\checked$}     & \CIRCLE \\
			NAVEX & $\checked$     & $\checked$     & $\checked$     & $\times$     & \multicolumn{1}{l|}{$\times$}     & \LEFTcircle  & $\times$     & $\checked$     & \multicolumn{1}{l|}{$\times$}     & \Circle     & $\times$     & $\checked$     & $\times$     & $\checked$     & $\times$     & $\checked$     & \multicolumn{1}{l|}{$\times$}     & \Circle & $\checked$     & $\times$     & $\times$     & $\checked$     & $\times$     & $\checked$     & $\checked$     & \multicolumn{1}{l|}{$\checked$}     & \LEFTcircle \\
			TCHECKER & $\checked$     & $\checked$     & $\checked$     & $\times$     & \multicolumn{1}{l|}{$\times$}     & \LEFTcircle  & $\checked$     & $\checked$     & \multicolumn{1}{l|}{$\checked$}     & \CIRCLE      & $\times$     & $\checked$     & $\times$     & $\checked$     & $\times$     & $\checked$     & \multicolumn{1}{l|}{$\times$}     & \Circle & $\times$     & $\checked$     & $\checked$     & $\checked$     & $\times$     & $\checked$     & $\checked$     & \multicolumn{1}{l|}{$\checked$}     & \LEFTcircle \\
			WHIP  & $\checked$     & $\checked$     & $\checked$     & $\times$     & \multicolumn{1}{l|}{$\times$}     & \LEFTcircle  & $\times$     & $\checked$     & \multicolumn{1}{l|}{$\times$}     & \Circle     & $\checked$     & $\checked$     & $\checked$     & $\checked$     & $\checked$     & $\checked$     & \multicolumn{1}{l|}{$\checked$}     & \CIRCLE & $\checked$     & $\times$     & $\checked$     & $\checked$     & $\checked$     & $\checked$     & $\checked$     & \multicolumn{1}{l|}{$\checked$}     & \LEFTcircle \\
			\textbf{Yama}  & $\checked$     & $\checked$     & $\checked$     & $\checked$     & \multicolumn{1}{l|}{$\checked$}     & \CIRCLE & $\checked$     & $\checked$     & \multicolumn{1}{l|}{$\checked$}     & \CIRCLE      & $\checked$     & $\checked$     & $\checked$     & $\checked$     & $\checked$     & $\checked$     & \multicolumn{1}{l|}{$\checked$}     & \CIRCLE & $\checked$     & $\checked$     & $\checked$     & $\checked$     & $\checked$     & $\checked$     & $\checked$     & \multicolumn{1}{l|}{$\checked$}     & \CIRCLE \\
			\hline
		\end{tabular}%
	}
	\label{yama_table_rq1}%
	 \begin{tablenotes}[flushleft]
		\scriptsize
		\item {$\checked$: True positive or true negative; $\times$: False positive or false negative; \Circle: Lack of ability; \LEFTcircle: Partial ability; \CIRCLE: Full ability; The superscripts \textit{N} denote a negative testcase.}
	\end{tablenotes}
\end{table*}%

\textbf{TCHECKER and NAVEX}. TCHECKER and NAVEX employ the same method for CFGs, utilizing php-ast\cite{nikic2023} to parse PHP source code into AST and PHPJoern\cite{malteskoruppa2020} to build CFGs and Data Dependency Graphs(DDGs). This results in similar analysis outcomes to some extent. As shown in Table \ref{yama_table_rq1}, TCHECKER possesses context-sensitive analysis capability, which NAVEX lacks, while their performance in the other three capability evaluations are nearly identical. Both TCHECKER and NAVEX perform poorly in the file inclusion analysis, with only three out of nine analysis tools lacking this capability, and TCHECKER and NAVEX are two of them. The function call analysis results indicate that TCHECKER lacks the capability to analyze taints returned by functions, as it incorrectly assesses the \textit{normal} and \textit{RnA}(Ret but not Arg, meaning the taint comes from the return) samples. NAVEX incorrectly assesses \textit{normal\_safe}(negative sample), \textit{AnR}(Arg but not Ret, meaning the taint comes from arguments), and \textit{RnA}, suggesting that NAVEX does not perform interprocedural analysis effectively and merely propagates taint information from the function parameters to the return values. Notably, because the original NAVEX code provided by the authors was unavailable, we used a fixed version(Navex\_fixed)\cite{m4yfly2020} for the experiments. We do not think that Navex\_fixed fully represents NAVEX's actual analysis capabilities.

\textbf{WHIP, Progpilot(P), and WAP}. WHIP collaborates with Progpilot and WAP, but the experimental results indicate that WHIP does not surpass the performance of the two collaborative tools, with results consistent with Progpilot(P). Progpilot(P) refers to executing in the specified analysis path mode, where Progpilot traverses the files within the analysis path. During the verification of the experimental results, it was unexpectedly discovered that executing Progpilot in the specified analysis file mode (analyzing only the specified file each time, referred to as Progpilot(F)) yields a $\CIRCLE$ rating in the function Call analysis. This indicates that Progpilot inherently possesses complete function call analysis capability but may achieve only a $\LEFTcircle$ rating in the Progpilot(P) mode due to implementation issues. Nevertheless, we choose to adopt the Progpilot(P) mode results because the specified path mode is more commonly used and is generally preferred for analyzing large-scale programs, whereas the specified file mode is not the primary choice. WAP performes the worst in the experiments, only achieving some success in the function call analysis. It demonstrates some interprocedural analysis capability when taints are present in the function parameters and return values, as it correctly identifies AnR and RnA.

\textbf{RIPS, Pixy, and PhpSafe}. RIPS and Pixy produce similar analysis results. They possess certain file inclusion analysis capabilities but cannot handle complex scenarios such as dynamic samples(\textit{dym}). Additionally, they perform poorly in context-sensitive analysis. For flow-sensitive analysis, both RIPS and Pixy incorrectly assess the \textit{normal\_safe} sample, and only two errors are observed. Upon further investigation, we find that these tools have errors when handling taint clearance logic within conditional statements, leading to these errors. PhpSafe aligns with RIPS and Pixy in the file inclusion and context-sensitive analysis but only possesses flow-insensitive analysis capability.

In summary, the lack of precise context sensitivity and interprocedural analysis capabilities resulted in poor performance of baselines in RQ1. Yama, with its precise parsing of opcodes, is able to satisfy the requirements of both function calls and file inclusion interprocedural analysis scenarios and achieve path-sensitive analysis, obtaining all correct results in the RQ1 experiment. This answers the first research question, indicating that Yama has basic data flow analysis capabilities.

\subsection{Complex Semantic Analysis Capability Assessment(RQ2)}

In this subsection, we evaluate the complex semantic analysis capabilities of Yama and baseline tools using dataset 2. The experimental results are presented in Table \ref{yama_table_rq2}, which illustrates the performance of each tool across three types of complex semantic. We focus on two detection metrics: false positive rate(FPR) and true positive rate(TPR). FPR indicates the proportion of non-vulnerable samples incorrectly identified as vulnerabilities by the detection tool, while TPR represents the proportion of vulnerable samples correctly identified. Due to space limitations, the evaluation results for each test case are provided in our repository\cite{Zhao2024}.

\begin{table}[htbp]
	\centering
	\caption{Evaluation of complex semantic analysis capability}
	\resizebox{0.5\textwidth}{!}{
		\begin{tabular}{c|cc|cc|cc|cc}
			\hline
			& \multicolumn{2}{c|}{C1\_type\_inference} & \multicolumn{2}{c|}{C2\_Dynamic\_features} & \multicolumn{2}{c|}{C3\_Built-in\_func} & \multicolumn{2}{c}{Total} \\
			\hline
			Tools & FPR   & TPR   & FPR   & TPR   & FPR   & TPR   & FPR   & TPR \\
			\hline
			WAP   & 12.5\% & 15.4\% & 0.0\% & 10.0\% & 0.0\% & 27.6\% & 9.1\% & 17.1\% \\
			Progpilot & 12.5\% & 36.5\% & 0.0\% & 20.0\% & 33.3\% & 37.9\% & 16.7\% & 32.4\% \\
			Pixy  & 0.0\% & 25.0\% & 0.0\% & 20.0\% & 33.3\% & 41.4\% & 8.3\% & 27.9\% \\
			RIPS  & 0.0\% & 32.7\% & 0.0\% & 46.7\% & 33.3\% & 65.5\% & 8.3\% & 45.0\% \\
			PhpSafe & 37.5\% & 19.2\% & 0.0\% & 13.3\% & 33.3\% & 27.6\% & 33.3\% & 19.8\% \\
			TCHECKER & 37.5\% & 25.0\% & 100.0\% & 23.3\% & 33.3\% & 34.5\% & 41.7\% & 27.0\% \\
			WHIP  & 25.0\% & 38.5\% & 0.0\% & 23.3\% & 33.3\% & 44.8\% & 25.0\% & 36.0\% \\
			\textbf{Yama}  & \textbf{0.0\%} & \textbf{98.1\%} & \textbf{0.0\%} & \textbf{100.0\%} & \textbf{0.0\%} & \textbf{100.0\%} & \textbf{0.0\%} & \textbf{99.1\%} \\
			\hline
		\end{tabular}%
	}
	\label{yama_table_rq2}%
\end{table}%

Overall, Yama achieved a TPR of 99.1\%, correctly identifying 110 true positives and 12 true negatives, with only one false negative and no false positives; thus, it is the top-performing tool. Following Yama, RIPS, WHIP, and Progpilot attain TPR of 45.0\%, 36.0\%, and 32.4\%, respectively, and they all maintain low FPRs, whereas the TPR of the other tools do not exceed 30\%. We further analyze each tool's performance within the three types of complex semantic.

\textbf{Type Inference (C1)}. In the challenge of type inference analysis, Yama outperforms all the other tools, with a TPR of 98.1\%. WHIP follow as the second-best analysis tool, achieving a TPR of 38.6\% and an FPR of 25\%. Progpilot and RIPS recorded TPR of 36.5\% and 32.7\%, respectively, while the TPR of the other tools do not exceed 30\%. To further analyze the reasons behind the underperformance of the baseline tools in C1, we categorize the C1 data into arrays, objects, and others based on the involvement of array or object variables. The experimental results of this categorization are presented in Table \ref{yama_table_rq2_ao}.

\begin{table}[htbp]
	\centering
	\caption{The results of array, object and others in C1}
		\resizebox{0.5\textwidth}{!}{
		\begin{tabular}{c|ccc|ccc|ccc}
			\hline
			& \multicolumn{3}{c|}{Array} & \multicolumn{3}{c|}{Object} & \multicolumn{3}{c}{Other} \\
			\hline
			Tools & True  & False & ACC   & True  & False & ACC   & True  & False & ACC \\
			\hline
			WAP   & 2     & 7     & 22.2\% & 4     & 23    & 14.8\% & 9     & 15    & 37.5\% \\
			Progpilot & 4     & 5     & 44.4\% & 10    & 17    & 37.0\% & 12    & 12    & 50.0\% \\
			Pixy  & 3     & 6     & 33.3\% & 3     & 24    & 11.1\% & 15    & 9     & 62.5\% \\
			RIPS  & 7     & 2     & 77.8\% & 5     & 22    & 18.5\% & 13    & 11    & 54.2\% \\
			PhpSafe & 4     & 5     & 44.4\% & 4     & 23    & 14.8\% & 7     & 17    & 29.2\% \\
			TCHECKER & 4     & 5     & 44.4\% & 7     & 20    & 25.9\% & 7     & 17    & 29.2\% \\
			WHIP  & 4     & 5     & 44.4\% & 10    & 17    & 37.0\% & 12    & 12    & 50.0\% \\
			\textbf{Yama}  & \textbf{9}     & \textbf{0}     & \textbf{100.0\%} & \textbf{27}    & \textbf{0}     & \textbf{100.0\%} & \textbf{23}    & \textbf{1}     & \textbf{95.8\%} \\
			\hline
		\end{tabular}%
		}
	\label{yama_table_rq2_ao}%
\end{table}%

From the table, we observe that in the array category, five baseline tools achieved accuracy rates(ACC) exceeding 44\%, with RIPS leading at 77.8\%. Similarly, in the other category, four baseline tools surpass 50\% ACC, with the highest reaching 62.5\%. However, in the object category, the top-performing baseline tool achieves only 37\% ACC. \textbf{This clearly indicates that type inference for object variables is a significant challenge affecting the performance of current popular static analysis tools}.

Additionally, a horizontal comparison reveals that four out of seven tools perform best in the other category, while the remaining three excel on the array category. \textbf{This suggests that although these tools handle the type inference relatively well for array variables, room for improvement remains}.

For Yama, we observe a false negative in the analysis of the other category. The test case combines JavaScript and PHP code, requiring the static analysis tool to interpret JavaScript logic, which exceeds Yama's current capabilities. Similarly, none of the other seven baseline tools are able to detect the vulnerability in this test case.

\textbf{Dynamic Features (C2)}. In addressing the challenge of dynamic feature analysis, Yama excels by detecting all vulnerabilities without any false positives, making it the top-performing tool. RIPS ranks second, achieving a TPR of 46.7\% with an FPR of 0\%. The TPR of the other tools does not exceed 25\%. To further investigate why the baseline tools underperform in C2, we categorize the C2 data into seven categories corresponding to the seven dynamic features (D1–D7) previously mentioned in subsection \ref{challenge2}. The results of this categorization experiment are presented in Table \ref{yama_table_rq2_dyn}. Notably, one test case from D2 and all test cases from D3 are categorized under both C2 and C3, because these test cases exhibit both dynamic features and characteristics of built-in functions.

\begin{table}[htbp]	
	\centering
	\caption{The results of different dynamic characteristics in C2}
	\resizebox{0.5\textwidth}{!}{
		\begin{tabular}{cccccccc}
			\hline
			\multirow{2}{*}{Tools} & D1    & D2*    & D3*    & D4    & D5    & D6    & D7 \\
			\cline{2-8}
			& \multicolumn{7}{c}{True(ACC)} \\
			\hline
			WAP   & 2(40.0\%) & 2(50.0\%) & 2(33.3\%) & 0(0.0\%) & 0(0.0\%) & 0(0.0\%) & 0(0.0\%) \\
			Progpilot & 2(40.0\%) & 3(75.0\%) & 2(33.3\%) & 0(0.0\%) & 0(0.0\%) & 0(0.0\%) & 2(22.2\%) \\
			Pixy  & 3(60.0\%) & 4(100.0\%) & 2(33.3\%) & 1(33.3\%) & 0(0.0\%) & 0(0.0\%) & 0(0.0\%) \\
			RIPS  & 3(60.0\%) & 3(75.0\%) & 2(33.3\%) & 2(66.7\%) & 3(50.0\%) & 3(60.0\%) & 2(22.2\%) \\
			PhpSafe & 2(40.0\%) & 0(0.0\%) & 2(33.3\%) & 1(33.3\%) & 1(16.7\%) & 0(0.0\%) & 1(11.1\%) \\
			TCHECKER & 3(60.0\%) & 0(0.0\%) & 2(33.3\%) & 0(0.0\%) & 1(16.7\%) & 2(40.0\%) & 1(11.1\%) \\
			WHIP  & 3(60.0\%) & 3(75.0\%) & 2(33.3\%) & 0(0.0\%) & 0(0.0\%) & 0(0.0\%) & 2(22.2\%) \\
			\textbf{Yama}  & \textbf{5(100.0\%)} & \textbf{4(100.0\%)} & \textbf{6(100.0\%)} & \textbf{3(100.0\%)} & \textbf{6(100.0\%)} & \textbf{5(100.0\%)} & \textbf{9(100.0\%)} \\
			\hline
			Avg ACC\dag & 51.4\% & 53.6\% & 33.3\% & 19.0\% & 11.9\% & 14.3\% & 12.7\% \\
			\hline
		\end{tabular}%
	}
	\label{yama_table_rq2_dyn}%
		 \begin{tablenotes}[flushleft]
		\scriptsize
		\item {*: One testcase in D2 and all testcases in D3 are categorized under both C2 and C3; \dag: Avg ACC without calculating Yama results}
	\end{tablenotes}
\end{table}%

From the table, we observe that Yama detected vulnerabilities across all types; thus, we excluded Yama when calculating the average accuracy to avoid skewing our analysis of the baseline tools. On average, baseline tools performed best in the D1 and D2 categories, followed by D3. A further analysis of the test cases in D1–D7 reveals that the five test cases in D1 do not involve function calls, and only one test case in D2 involves a function call. In contrast, all test cases in D3–D7 involve function calls or object method calls. \textbf{This indicates that dynamic features involving function calls and object method calls present significant challenges for baseline tools to analyze correctly}. This finding aligns with the conclusions in RQ1, which highlighted that the function call analysis capabilities of baseline tools need improvement and that type inference for object variables is a significant factor affecting current static analysis tools, as discussed in C1. The 33.3\% ACC achieved by baseline tools in D3 primarily from correctly identifying two negative samples.

\textbf{Built-in Functions (C3)}. As shown in Table \ref{yama_table_rq2}, Yama outperforms all other tools in the built-in function analysis by detecting all the vulnerabilities without any false positives. RIPS ranks second with a TPR of 65.5\% and an FPR of 33.3\%. WHIP and Pixy followed, achieving TPR of 44.8\% and 44.4\%, respectively. The remaining tools do not surpass a 40\% TPR.

RIPS's strong performance is attributed to its comprehensive modeling of 952 built-in functions, which enhances its analysis capabilities when meeting built-in functions. WHIP's results are based on re-analyzing the merged data flows from Progpilot and WAP, with Progpilot--having a TPR of 37.9\% likely being the primary contributor to WHIP's effectiveness. Progpilot models 415 built-in functions, which explains why both it and WHIP produce relatively good analysis outcomes.

Pixy generalizes all the PHP built-in functions available as of 2005 and assumes that all the built-in function return values carry taints. While this approach leads to over-tainting (since many built-in function return values do not actually carry taints), it allowed Pixy to achieve a TPR of 44.4\% in C3. Additionally, TCHECKER achieves a TPR of 34.5\%, having modeled a limited number of built-in functions.

\textbf{It is evident that both the quantity and accuracy of built-in function modeling are crucial for effectively addressing the analysis challenges posed by built-in functions}. Yama models 1,097 built-in functions and introduced concrete execution techniques to resolve issues that cannot be addressed through function modeling alone. Furthermore, Yama's foundational data flow analysis capabilities (RQ1) and type inference capabilities (RQ2-C1) are critical for overcoming the built-in function challenges in C3. Deficiencies in these two capabilities may lead to inaccurate taint propagation paths in data flow analysis. Additionally, a lack of built-in function analysis capabilities can result in interruptions of taint propagation during function calls and taint sinks.

In summary, type inference, dynamic features, and built-in functions pose significant challenges to baseline tools. The experimental results answered the second research question: Yama can effectively address the three current challenges in PHP static analysis. This is because the semantics expressed by opcodes are precise, and we have a comprehensive and systematic analysis of opcodes.

\subsection{Vulnerability Detection Capability Assessment (RQ3)}

In this subsection, we evaluate the vulnerability detection capabilities of Yama and the baseline tools on real-world applications using the dataset 3. The experimental results are presented in Table \ref{yama_rq3}. We use TP and FP to denote the numbers of true positives and false positives, respectively. Overall, Yama detects 109 true positive vulnerabilities and 78 false positives, achieving an ACC of 58.3\% and outperforming all other tools. Additionally, we assessed Yama's ability to discover new vulnerabilities, as indicated by NV in Table \ref{yama_rq3}. Yama reports 38 new vulnerabilities, including 6 RCE, 2 SQLI, 2 PT, and 28 XSS vulnerabilities. Specifically, the 2 SQLI vulnerabilities allow attackers to write web shells to the server, and the 2 PT vulnerabilities enable arbitrary file inclusion, effectively leading to RCE vulnerabilities. The XSS vulnerabilities permit non-privileged users to steal session cookies. We have responsibly disclosed all our findings to the relevant vendors. At the time of writing, all 38 vulnerabilities have been confirmed, including 34 new CVEs\cite{Zhao2024}. 

\begin{table*}[htbp]
	\centering
	\caption{Evaluation results of vulnerability detection.}
	\resizebox{\textwidth}{!}{
		\begin{tabular}{cccccccccccccccccc}
			\hline
			Application & TP$^Y$ & FP$^Y$ & TP$^W$ & FP$^W$ & TP$^P$ & FP$^P$ & TP$^R$ & FP$^R$ & TP$^{PH}$ & FP$^{PH}$ & TP$^{WA}$ & FP$^{WA}$ & TP$^T$ & FP$^T$ & UP    & UN    & NV \\
			\cmidrule(r){1-1} \cmidrule(r){2-3} \cmidrule(r){4-5}  \cmidrule(r){6-7}  \cmidrule(r){8-9}  \cmidrule(r){10-11} \cmidrule(r){12-13} \cmidrule(r){14-15}  \cmidrule(r){16-18}
			wordpress (6.6) & 0     & 0     & -     & -     & 0     & 0     & -     & -     & 0     & 182   & 0     & 2     & -     & -     & 0     & 0     & 0 \\
			DVWA (1.9) & 28    & 6     & 18    & 9     & 18    & 9     & 18    & 7     & 6     & 13    & 9     & 5     & 6     & 3     & 10     & 0     & 0 \\
			dolibarr (12.0.0) & 1     & 0     & -     & -     & 0     & 0     & -     & -     & 1     & 3371  & 0     & 394   & -     & -     & 0     & 0     & 0 \\
			organizr (1.9) & 3     & 0     & 2     & 37    & 2     & 167   & 1     & 78    & 0     & 12    & 2     & 4     & 1     & 10    & 1     & 0     & 3 \\
			Joomla (5.1.2) & 0     & 0     & 0     & 15    & 0     & 17    & -     & -     & 0     & 1365  & 0     & 0     & -     & -     & 0     & 0     & 0 \\
			SuiteCRM (7.12.6) & 6     & 10    & 4     & 388   & 4     & 388   & -     & -     & 0     & 1088  & 0     & 0     & -     & -     & 2     & 0     & 0 \\
			leantime (2.1.5) & 3     & 0     & 3     & 70    & 3     & 92    & 3     & 27    & 2     & 52    & 0     & 20    & 0     & 13    & 0     & 0     & 0 \\
			glpi (10.0.16) & 4     & 2     & -     & -     & 1     & 28    & 0     & 205   & 4     & 1004  & 0     & 2     & -     & -     & 0     & 0     & 4 \\
			dzzoffice (2.02.1) & 1     & 2     & 1     & 101   & 1     & 115   & 1     & 135   & 0     & 72    & 0     & 7     & 0     & 11    & 0     & 0     & 1 \\
			librenms (21.1.0) & 4     & 1     & 4     & 92    & 1     & 110   & 4     & 82    & 0     & 141   & -     & -     & -     & -     & 0     & 0     & 0 \\
			microweber (2.0.16) & 2     & 2     & 2     & 44    & 2     & 44    & 2     & 85    & 2     & 241   & 1     & 10    & -     & -     & 0     & 0     & 2 \\
			microweber (1.2.3) & 15    & 1     & 7     & 46    & 7     & 46    & 7     & 74    & 4     & 140   & 4     & 4     & -     & -     & 8     & 0     & 0 \\
			openemr (6\_0\_0) & 1     & 4     & 1     & 480   & 1     & 515   & -     & -     & 1     & 1304  & 0     & 28    & -     & -     & 0     & 0     & 0 \\
			shopware (6.4.4.0) & 1     & 0     & -     & -     & 0     & 0     & 0     & 3     & 1     & 5815  & -     & -     & -     & -     & 0     & 0     & 0 \\
			phpipam (1.6) & 7     & 3     & -     & -     & 4     & 327   & -     & -     & 3     & 946   & 0     & 0     & 0     & 2     & 3     & 0     & 7 \\
			WDScanner (latest) & 1     & 8     & 0     & 8     & 0     & 11    & 0     & 27    & 0     & 7     & 0     & 10    & 0     & 0     & 1     & 0     & 0 \\
			bjyadmin (latest) & 2     & 0     & -     & -     & 0     & 0     & 2     & 42    & 1     & 46    & 1     & 2     & 1     & 2     & 0     & 0     & 2 \\
			Gazelle (latest) & 7     & 31    & 4     & 410   & 4     & 410   & 4     & 480   & 4     & 640   & 0     & 519   & 0     & 284   & 3     & 0     & 3 \\
			phpbb (3.3.10) & 1     & 0     & -     & -     & 0     & 0     & 1     & 15    & 0     & 46    & 0     & 2     & 0     & 2     & 0     & 0     & 0 \\
			unmark (1.9.2) & 1     & 1     & 1     & 22    & 1     & 26    & 1     & 4     & 1     & 10    & 1     & 0     & -     & -     & 0     & 0     & 1 \\
			icecoder (8.1) & 4     & 4     & 2     & 93    & 2     & 96    & 1     & 206   & 3     & 130   & 1     & 21    & 0     & 146   & 1     & 0     & 3 \\
			openflights (latest) & 6     & 3     & 4     & 130   & 4     & 139   & 4     & 54    & 4     & 41    & 0     & 1     & 4     & 32    & 2     & 0     & 4 \\
			RPI-Jukebox (2.7.0) & 8     & 0     & 3     & 86    & 3     & 86    & 2     & 126   & 3     & 131   & 2     & 68    & 1     & 13    & 5     & 0     & 6 \\
			PicUploader (latest) & 3     & 0     & -     & -     & 2     & 5     & 2     & 75    & 2     & 209   & -     & -     & -     & -     & 1     & 0     & 2 \\
			\cmidrule(r){1-1} \cmidrule(r){2-3} \cmidrule(r){4-5}  \cmidrule(r){6-7}  \cmidrule(r){8-9}  \cmidrule(r){10-11} \cmidrule(r){12-13} \cmidrule(r){14-15}  \cmidrule(r){16-18}
			\textbf{Total} & \textbf{109}   & \textbf{78}    & 56    & 2031  & 60    & 2631  & 52    & 1723  & 42    & 17006 & 21    & 1099  & 13    & 518   & 37    & 0     & 38 \\
			\hline
		\end{tabular}%
	}
	\label{yama_rq3}%
	\begin{tablenotes}[flushleft]
		\scriptsize
		\item {The superscripts $Y,W,P,R,PH,WA,T$ denote the results of Yama, WHIP, Progpilot, RIPS, PhpSafe, WAP, and TCHECKER, respectively. \textbf{NP} denotes the number of vulnerabilities found by Yama but not by the other tools. \textbf{UN} denotes the number of vulnerabilities found by the other tools but not by Yama. \textbf{NV} denotes the number of new vulnerabilities found by Yama. The \textbf{hyphen (-)} indicates that the tool cannot output results within 6 hours or report errors during detection.}
	\end{tablenotes}
\end{table*}%

In Table \ref{yama_rq3}, we use superscripts Y, W, P, R, PH, WA, and T to represent the results of Yama, WHIP, Progpilot, RIPS, PhpSafe, WAP, and TCHECKER, respectively. UP denotes the number of vulnerabilities found only by Yama, while UN represents the number of vulnerabilities not detected by Yama but found by other tools. Overall, Yama's vulnerability detection capability significantly surpasses that of the other tools, identifying all vulnerabilities reported by others (i.e., UN=0) and discovers 37 additional vulnerabilities that other tools failed to detect (i.e., UP=37).

\subsubsection{Comparison with Related Works}

Below, we compare Yama with related works and discuss how the different design choices affect the detection results.

\textbf{WHIP, Progpilot, and WAP}. Progpilot detects 60 true positive vulnerabilities, making it the best-performing baseline tool. However, compared with Yama, it still misses 49 vulnerabilities. Among these 49 vulnerabilities, 20 involve tainted variables propagated back and forth across multiple procedures, and 9 vulnerabilities depend on the complex propagation of object properties for taint tracking. Yama's accurate type inference and interprocedural analysis support the analysis of these vulnerabilities, whereas Progpilot does not. Seven vulnerabilities rely on path-sensitive data flow analysis; these vulnerabilities exist within one of the branches. Progpilot, lacking path-sensitive data flow analysis capabilities, defaults to analyzing a single branch, thereby failing to analyze the vulnerable branch. Additionally, six vulnerabilities employ reversible sanitization operations. Yama supports the analysis of reversible sanitization operations, correctly restoring taint attributes after such operations, while Progpilot does not. Seven vulnerabilities depend on the precise parsing of built-in functions beyond Progpilot's modelling scope, while Yama's dangerous parameter list and concrete execution can support the analysis of these vulnerabilities.

WAP, equipped with a false positive prediction mechanism that uses a logistic regression algorithm to verify the authenticity of vulnerabilities, achieved the second-lowest number of false positives among the baseline tools. However, due to its limited interprocedural analysis capabilities, WAP is only able to detect 21 true positive vulnerabilities. WHIP does not achieve results surpassing those of the two collaborative tools, and it encounters errors when analyzing eight real-world applications, preventing us from obtaining experimental results for WHIP in these applications. In the table, a hyphen (-) is used to indicate detection errors or cases where results are not obtained within six hours. It can be observed that WAP experiences issues when analyzing three real-world applications.

\textbf{RIPS, PhpSAFE, TCHECKER, and Pixy}. RIPS detects 52 true positive vulnerabilities, 57 less than to Yama. This shortfall occurs primarily because RIPS performs data flow analysis based on ASTs, making comprehensive and accurate type inference and program state storage challenging,  while opcode-based Yama does not face these challenges. Additionally, RIPS struggles when analyzing the taint propagation trends of complex object properties and cannot handle taint propagation in dynamic; these are key reasons for its failure to detect the remaining vulnerabilities. PhpSAFE detected only 42 true positive vulnerabilities; besides sharing RIPS's limitations, PhpSAFE is implemented to support single-file detection only, rendering it incapable of handling multiple interprocedural taint propagations. This limitation accounts for the 10 fewer true positives detected compared to RIPS.

TCHECKER, despite performing well in RQ1 and RQ2, detects only 13 true positive vulnerabilities. This is mainly due to implementation issues: in 24 real-world applications, TCHECKER fails to analyze 12 vulnerabilities (50\%), severely impacting its overall results. Pixy does not conduct a complete vulnerability mining experiment on real applications because, upon analysis, we find that Pixy reports syntax errors in over 80\% of the files. This is attributable to Pixy's design for PHP4 syntax, indicating its inability to analyze the current versions of real-world applications.

In summary, Yama's precise context-sensitive and path-sensitive interprocedural taint analysis based on precise parsing of opcodes semantics is the core of outperforming baseline work, where reversible taint sanitization, dangerous parameters list, and concrete execution are also key to helping Yama detect more vulnerabilities in real-world applications.

\subsubsection{False Positives}

Yama reports a total of 78 false positives, attributable to three primary reasons:

\textbf{Coarse Granularity Taint Sanitization}. Although Yama endeavors to cover all known taint sanitization operations, its recognition and differentiation of sanitization functions for specific vulnerability types remain coarse-grained. For example, the function mysql\_real\_escape\_string effectively defends against character-based SQL injection attacks but is ineffective against numeric SQL injections, which typically do not require special characters such as quotes. Yama does not further distinguish between different cases within the same vulnerability type, leading to false positives due to the lack of fine-grained differentiation of sanitization functions.

\textbf{Regular Expression-Based Taint Sanitization}. Some applications use regular expression functions to validate data formats, such as using preg\_match to check for the presence of characters including ., /, or \textbackslash. This method can sanitize taints for path traversal vulnerabilities, which exploit these characters for directory traversal, but it does not sanitize taints for other vulnerability types. Yama does not differentiate the key characters required by different vulnerabilities, resulting in false positives when it cannot accurately determine the sanitization effectiveness based on the current vulnerability type.

\textbf{Lack of Constraint Solving Capabilities}. In practice, we observe that conditional constraints along taint propagation paths in applications can conflict with the conditions required for successfully exploiting vulnerabilities. Yama, lacking constraint-solving capabilities, may generate false positives in such scenarios. For example, Yama falsely reports a file upload vulnerability where the application checks if the uploaded file's extension satisfies certain constraints (e.g., it must be jpg, jpeg, or png). Since the uploaded filename is tainted, Yama assumes that the condition can be satisfied. Upon meeting this constraint, the taint propagates to the sink function, and the file is uploaded. However, due to the prior constraint, only files with the specified extensions can be uploaded, which does not satisfy the exploitation conditions of a file upload vulnerability.

\subsubsection{False Negatives}

Although Yama detects all vulnerabilities reported by other tools, it still exhibits some false negative detections, primarily due to encountering vulnerabilities beyond its current analytical scope. For example, Yama cannot detect stored XSS vulnerabilities because it lacks modeling for the data flows of taint variables within databases. Additionally, Yama cannot analyze the invocation relationships of PHP files within JavaScript code, leading to loss of data flow and subsequent missed detections. Yama is highly extensible, supporting the addition of new analysis algorithms as plugins and enabling further analysis on existing data flow results. We plan to expand Yama's vulnerability analysis capabilities in future work.

\subsubsection{Performance}

In Table \ref{yama_rq3_time}, we use superscripts Y, W, P, R, PH, WA, and T to denote the analysis time required by each tool for each application. As before, a hyphen (-) indicates an instances where the tool crashes or fails to produce results within 6 hours. WAP has the lowest time overhead among all tools, with a total analysis time of 653 seconds for 21 applications. Next is RIPS, with a total analysis time of 1,054 seconds for 18 applications; however, RIPS encounters issues when analyzing 6 applications, reducing its total analysis time. TCHECKER completes the analysis on only 12 applications, totaling 1,187 seconds.

\begin{table}[htbp]
	\centering
	\caption{Performance of each detection method.}
        \resizebox{0.5\textwidth}{!}{
	\begin{tabular}{|c|r|r|r|r|r|r|r|}
		\hline
		Application & Time$^Y$ & Time$^W$ & Time$^P$ & Time$^R$ & Time$^{PH}$ & Time$^{WA}$ & Time$^T$ \\
		\hline
		wordpress (6.6) & 100s  & -     & 433s  & -     & 198s  & 21s   & - \\
		DVWA (1.9) & 23s   & 494s  & 116s  & 11s   & 11s   & 11s   & 29s \\
		dolibarr (12.0.0) & 426s  & -     & 3178s & -     & 802s  & 158s  & - \\
		organizr (1.9) & 116s  & 37s   & 11s   & 10s   & 156s  & 5s    & 14s \\
		Joomla (5.1.2) & 93s   & 8363s & 1277s & -     & 1084s & 16s   & - \\
		SuiteCRM (7.12.6) & 108s  & 4388s & 1143s & -     & 1293s & 22s   & - \\
		leantime (2.1.5) & 6s    & 190s  & 61s   & 7s    & 27s   & 3s    & 14s \\
		glpi (10.0.16) & 182s  & -     & 962s  & 249s  & 1124s & 141s  & - \\
		dzzoffice (2.02.1) & 16s   & 1655s & 541s  & 26s   & 40s   & 17s   & 61s \\
		librenms (21.1.0) & 41s   & 4979s & 729s  & 93s   & 207s  & -     & - \\
		microweber (2.0.16) & 99s   & 3173s & 1264s & 60s   & 341s  & 13s   & - \\
		microweber (1.2.3) & 35s   & 2212s & 782s  & 44s   & 238s  & 14s   & - \\
		openemr (6\_0\_0) & 204s  & 6603s & 2044s & -     & 409s  & 44s   & - \\
		shopware (6.4.4.0) & 44s   & -     & 2126s & 64s   & 515s  & -     & - \\
		phpipam (1.6) & 120s  & -     & 4612s & -     & 117s  & 16s   & 65s \\
		WDScanner (latest) & 1s    & 9s    & 2s    & 1s    & 2s    & 3s    & 3s \\
		bjyadmin (latest) & 21s   & -     & 889s  & 234s  & 154s  & 123s  & 610s \\
		Gazelle (latest) & 11s   & 40s   & 10s   & 12s   & 20s   & 10s   & 32s \\
		phpbb (3.3.10) & 80s   & -     & 151s  & 31s   & 218s  & 16s   & 216s \\
		unmark (1.9.2) & 5s    & 410s  & 132s  & 5s    & 36s   & 9s    & - \\
		icecoder (8.1) & 74s   & 79s   & 25s   & 6s    & 36s   & 3s    & 136s \\
		openflights (latest) & 6s    & 16s   & 4s    & 3s    & 12s   & 3s    & 5s \\
		RPI-Jukebox (2.7.0) & 4s    & 22s   & 7s    & 2s    & 10s   & 5s    & 4s \\
		PicUploader (latest) & 50s   & -     & 13s   & 197s  & 1264s & -     & - \\
		\hline
		Total & 1865s & 32670s & 20515s & 1054s & 8313s & 653s  & 1187s \\
		\hline
	\end{tabular}%
        }
	\label{yama_rq3_time}%
	\begin{tablenotes}[flushleft]
		\scriptsize
		\item {The superscripts $Y,W,P,R,PH,WA,T$ denote the results of Yama, WHIP, Progpilot, RIPS, PhpSafe, WAP, and TCHECKER, respectively. The \textbf{hyphen (-)} indicates that the tool cannot output results within 6 hours or report errors during detection.}
	\end{tablenotes}
\end{table}%

Yama successfully analyzes all 24 applications, with a total analysis time of 1,865 seconds. Subtracting the analysis times for the 6 applications that RIPS fails to analyze, Yama's total analysis time for the remaining 18 applications is 814 seconds, which is actually less than that of RIPS. Following in order of increasing analysis time are PhpSafe, Progpilot, and WHIP. Progpilot, capable of analyzing all 24 applications and detecting the most true positives among the baselines, incurred a total analysis time of 20,515 seconds--11 times longer than Yama. WHIP exhibited the longest analysis time among all tools, as it requires iterative analyses using two tools until no new data flows emerge, resulting in a total analysis time of 32,670 seconds for 16 applications.

\section{Limitations and Future Work}

This section describes the current limitations of Yama and our plans to handle them as part of our future work. 

\textbf{Taint Sanitization}. Currently, our taint sanitization approach is generic and does not specify operations for each vulnerability type. We plan to develop an algorithm to abstract semantic constraints for payloads, enabling generalized taint sanitization recognition.

\textbf{Constraint Solving}. Yama lacks constraint-solving abilities, leading to potential false positives. We aim to introduce a lightweight constraint solver to check if path condition constraints conflict with payload semantics, balancing accuracy and performance.

\textbf{Vulnerability Type Extension}. The types of vulnerabilities that Yama can currently detect are limited, unable to identify vulnerabilities such as stored XSS and second-order SQL injection vulnerabilities. We plan to expand Yama's scope by establishing threat models for additional vulnerabilities.

\textbf{Automatic Vulnerability Verification}. Yama reports vulnerabilities through static analysis, requiring manual verification. To improve this process, we plan to automate verification, potentially using targeted fuzzing based on static analysis findings.

\section{Related Work}

In recent years, detecting vulnerabilities in web applications has garnered significant attention. Previous research has focused primarily on two methods: static analysis and dynamic testing. For static analysis,  data flow (taint) analysis has emerged as the predominant research approach. Tools such as TCHECKER, NAVEX, RIPS, Progpilot, and PhpSAFE have performed extensive analyses at the AST level. However, an AST emphasizes showcasing the code's structure rather than the program's execution process, posing inherent challenges in representing program control flow and semantics. This exacerbates the susceptibility of static analysis methods--which are already labor-intensive and prone to implementation errors--to further imprecise\cite{Shi2024}. Notably, WHIP attempts to enhance analysis capabilities by sharing data flows between two static analysis methods. However, this approach is highly dependent on the precise data flow analysis in the collaborating tools; the sharing of data flows introduces both precise and imprecise data flow, further amplifying the false-positive issues existing in static analysis.

Dynamic testing, on the other hand, can identify application issues within the actual execution environment\cite{Trickel2023}. However, the complex business logic of web applications renders web pages increasingly dynamic and intricate, posing challenges for dynamic testing methods to achieve ideal code coverage and to effectively access deeply embedded code within web applications. Furthermore, the mutation feedback mechanism of test cases in dynamic testing is not intelligent enough to satisfy the format requirements of payloads for web application vulnerabilities, making it difficult to produce optimal testing results. Hybrid methods attempt to combine the advantages of static and dynamic approaches\cite{Alhuzali2018}, but effectively utilizing information from static analysis to assist dynamic testing remains an insurmountable challenge in current research.

\section{Conclusion}

In this paper, we have presented Yama, a context-sensitive and path-sensitive interprocedural data flow analysis method for PHP taint-style vulnerability detection, improving detection accuracy. Our approach is based on the observation that PHP opcodes have precise semantics and clear control flow, which can make data flow analysis more precise and efficient. Utilizing this insight, we have established parsing rules for PHP opcodes and presented a precise and efficient analysis method. Extensive experimental evaluations have been conducted, demonstrating that Yama can precisely infer variable types, handle PHP's complex dynamic features, and mitigate the problem of built-in function analysis. Yama has successfully discovered and reported 38 zero-day vulnerabilities in 24 GitHub projects with over 1,000 stars, and 34 new CVE IDs have been assigned. The experimental results demonstrate that Yama can effectively detect taint-type vulnerabilities in modern PHP applications. We hope this work lays the foundation for further exploration of the application of PHP Opcodes in PHP data flow analysis.

\bibliographystyle{IEEEtran}
\bibliography{ref.bib}

\newpage

 




\vfill

\end{document}